\documentclass[apj]{emulateapj}

\usepackage{times}

\shorttitle{}                    
\shortauthors{Brooks}

\shorttitle{Properties of the diffuse background}

\begin{document}


\title{Properties of the diffuse emission around warm loops in solar active regions}

\author{David H. Brooks\altaffilmark{1,2}}

\affiliation{\altaffilmark{1}College of Science, George Mason University, 4400 University Drive,
  Fairfax, VA 22030}

\altaffiltext{2}{Current address: Hinode Team, ISAS/JAXA, 3-1-1 Yoshinodai, Chuo-ku, Sagamihara,
  Kanagawa 252-5210, Japan}


\begin{abstract}
Coronal loops in active regions are the subjects of intensive investigation, but the important diffuse 'unresolved'
emission in which they are embedded has received relatively little attention.
Here we measure the densities and emission measure
(EM) distributions
of a sample of background-foreground regions surrounding warm (2\,MK) coronal loops, and introduce two new aspects to the analysis.
First, we infer the EM distributions only from temperatures that
contribute to the same background emission. Second, we measure the background emission co-spatially
with the loops so that the results are truly representative of the immediate loop environment. 
The second aspect also allows us to take advantage of the presence of embedded loops to infer information
about the (unresolvable) magnetic field in the background. We find that 
about half the regions in our sample have narrow
but not quite isothermal EM distributions 
with a peak temperature of 1.4--2\,MK. The other half have broad EM distributions
(Gaussian width $>$3$\times$10$^{5}$\,K), and the width of the EM appears to be correlated with peak temperature. Densities in
the diffuse background are $\log$ (n/cm$^{-3}$) = 8.5-9.0.
Significantly, these densities and temperatures imply that the co-spatial background 
is broadly comptabile with static equilibrium theory (RTV scaling laws) provided the unresolved field
length is comparable to the embedded loop length. For this agreement to break down, the field length
in most cases would have to be substantially longer than the loop length, a factor of 2--3 on average,
which for our sample approaches the dimensions of only the largest active regions.
\end{abstract}



\section{introduction}
\label{introduction}

Solar active regions typically comprise a hot ($\sim$4\,MK) core, surrounded 
by higher lying warm ($\sim$1.7\,MK) loops, and peripheral cooler fan structures ($\sim$1\,MK). The whole
region is embedded in diffuse unresolved emission from the foreground and background around loops. 
Understanding the physical properties of these structures is key to providing constraints 
on theories of how they are heated. 

There is a vast literature on measurements of the temperatures and densities of coronal loops. \citet{reale_2014} 
gives a very comprehensive review. One specific issue that has attracted a great deal of attention over the years
is whether coronal loops are consistent with static equilibrium theory and therefore whether they are heated 
steadily or impulsively. Observations of high temperature soft X-ray loops from the {\it Yohkoh} satellite by
\citet{porter&klimchuk_1995} appeared to show that these loops last longer than the expected radiative and conductive cooling time and
that they are therefore heated in a quasi-steady fashion. 
Measurements of the temperatures, pressures, and lengths 
also showed that the loops are consistent with the static equilibrium scaling laws 
derived by \citet{rosner_etal1978}. 
Consistency with static equilibrium theory is now thought to indicate that loops are 
heated quasi-steadily by high frequency impulsive heating \citep{klimchuk_2006}. 
Measurements of intensities, Doppler, and non-thermal
velocities at the footpoints of active region core loops from the {\it Hinode} EUV Imaging Spectrometer \citep[EIS,][]{culhane_etal2007a}
also do not show much variation over many hours \citep{brooks&warren_2009}, and the emission measure (EM) distributions at
the loop apex are often narrow with a steep slope
\citep{warren_etal2011b,winebarger_etal2011}. These results imply that they are not evolving through a broad range of temperatures \citep{warren_etal2012},
though there is evidence that the EM slope varies within a single AR \citep{delzanna_etal2015}, between others \citep{tripathi_etal2011}, and as a function
of AR age \citep{ugarteurra&warren_2012}. Furthermore, intensity time-lag analysis shows that these loops 
are partially cooling and being re-heated at high temperatures \citep{viall&klimchuk_2017}.
Higher spatial and temporal resolution observations also detect variability on shorter time-scales \citep{testa_etal2013}, though
this may depend on the initial conditions in the heated loop \citep{polito_etal2018}.

The warm loops formed at lower temperatures also persist longer than expected cooling times \citep{lenz_etal1999},
and the majority have narrow, near isothermal temperature distributions \citep{delzanna_2003,aschwanden&nightingale_2005,warren_etal2008a}. 
In contrast to the hot core
loops, however, the warm loops are clearly over-dense compared to static equilibrium theory \citep{lenz_etal1999,aschwanden_etal2001,winebarger_etal2003b}.

It was recognized early on that emission from the background/foreground complicates the analysis of coronal loops
\citep{klimchuk_etal1992}.
\citet{delzanna&mason_2003} argue that background subtraction is essential for obtaining accurate measurements.
This is because the background emission itself can account for $>$70\% of the intensity measured along the line-of-sight 
towards the loop \citep{delzanna&mason_2003,aschwanden&nightingale_2005,brooks_etal2012}, though this may be reduced
in higher spatial resolution observations \citep{delzanna_2013b}. Such a large contribution clearly has a significant
potential impact on the derived properties of loops. For one, when determing the thermal disttribution, 
it makes it difficult to determine whether weak emission
at the low- and high-temperature extremes is actually coming from the same loop structure. This is often handled by 
only including the emission from spectral lines with cross-loop intensity profiles that are correlated \citep{warren_etal2008a}. 
Background removal techniques, however, and the locations where the background
is measured, also have a significant influence on results. \citet{aschwanden_etal2008} report that the thermal width of warm loops 
varies with the proximity of the loop to the location where the background is measured. 
\citet{schmelz_etal2001} point out that different methods of background 
subtraction can even result in different conclusions on the location of heating in the same loop \citep{priest_etal1998,aschwanden_2001,reale_2002}.
For the higher temperature core loops, background subtraction
reduces the amount of lower temperature emission, leading to steeper EM slopes \citep{delzanna_etal2015}.

Despite the widespread prevalence and large contribution of this diffuse emission, and therefore its importance in understanding
active region and coronal heating, the properties of the background itself have been relatively less studied than those of coronal loops, and those
detailed studies that have been made have tended to focus on only a few examples.
\citet{delzanna&mason_2003} measured EM distributions in the diffuse background adjacent to
and within about 10$''$ or so of
the leg of a 0.7--1.1\,MK isothermal loop observed by the {\it SOHO} Coronal Diagnostic Spectrometer \citep[CDS,][]{harrison_etal1995}.
They found fairly flat EM distributions (not isothermal) in the 1--2\,MK range and electron densities below $\log$ (n/cm$^{-3}$) = 9.
\citet{cirtain_etal2006} examined radial intensity profiles in the background emission above three ARs observed at the limb by 
CDS and TRACE \citep[Transition Regions And Coronal Explorer,][]{handy_etal1999}. They found that for the quiescent ARs they studied,
the off-limb radial intensity profiles were consistent with an isothermal hydrostatic corona. In contrast, \citet{delzanna_2012} found
a multi-thermal EM distribution peaking at $\sim$2\,MK for an area of background emission about 50$''$ above the limb and close to an AR observed by EIS.
\citet{subramanian_etal2014} also used EIS data to examine the diffuse emission at $\sim$2.2\,MK a few tens of arcseconds above two limb active regions in an 
attempt to clearly isolate it from 
the hot loops at these temperatures. They found peak temperatures of 1.8\,MK and also derived densities below $\log$ (n/cm$^{-3}$) = 9
for all the regions they analyzed. Taking a different approach, \citet{viall&klimchuk_2011} applied their novel intensity time-lag analysis
to observations of a location of diffuse emission in an AR core observed by
the Atmospheric Imaging Assembly \citep[AIA,][]{lemen_etal2012} on the {\it Solar
Dynamics Observatory} \citep[SDO,][]{pesnell_etal2012},
and concluded that it was consistent with a long duration storm of low-frequency
(impulsive) heating. Later, \citet{viall&klimchuk_2012} studied the diffuse emission in a whole AR and found evidence of widespread 
cooling and evolution of the plasma. 
\citet{delzanna_2013b} also has an interesting discussion of the effects of the increased spatial resolution of AIA on the background
intensities measured by EIS. He showed that an almost factor of 2 increase in background emission in the core of an AR is likely 
due to its lower spatial resolution.

Given what we now know about the analysis of coronal loops, there are a number of important issues that should be addressed
before we can reliably conclude what the properties of the background emission are. \citet{delzanna_etal2015} point out that the slope
of the EM distribution for hot loops in the AR core is increased when the background emission is removed, and, as
discussed above, we know that the width of the thermal distribution for warm loops is strongly affected
by how close to the loop the background is measured \citep{aschwanden_etal2008}. We also know that it is difficult to 
incorporate weak emission at temperatures far removed from
the peak temperature and it is not always clear whether it is coming from the same structure and should be included or excluded.
This clearly affects the width of the
thermal distribution. So a focused analysis of the background emission should consider two important questions.

First, are the properties of the background
emission measured adjacent to loops, or of diffuse emission high above and remote from loops, representative of the background in which 
the loop itself is embedded? Referring again to the discussion by \citet{delzanna_2013b}, the increase in background emission in
the AR core compared to the periphery (about 80$''$ away), suggested as being due to the lower spatial resolution of EIS compared to
AIA, is much larger than what is seen over the much shorter distances (a few arcsecs) across the loops themselves (see \citet{delzanna_2013b} Fig. 8).

Second, does including the emission from spectral lines at all temperatures result in a truly representative EM distribution for 
the background emission? 

In this paper we attempt to address these questions by developing coronal loop analysis techniques to be applied to the background emission. 
First, we measure the diffuse emission simultaneously and co-spatially with a sample of 
warm coronal loops, so that it is truly representative of the background in which they are embedded. Second, we introduce a technique that
examines the spatial correlation between background intensities in different spectral lines to 
determine whether the emission at different temperatures is coming from the same structures. 

Our analysis and results refine those of 
previous studies and allow further potentially interesting insights. 
For example, the densities reported in the background emission are somewhat lower than found
in warm loops \citep[$\log$ (n/cm$^{-3}$) $>$ 9,][]{brooks_etal2012}, so there is a question as to whether the diffuse emission could be compatible with static equilibrium theory?
\citet{subramanian_etal2014} discuss the slopes of the EM distributions
in the diffuse emission regions they analyzed and find that they are consistent with both low- and high-frequency heating, and show
characteristics
that are broadly similar to those of the high temperature core loops. It is difficult to make a direct comparison with loop scaling laws,
however, because the lengths of the field lines cannot be 
determined in the unresolved emission.
By measuring the background co-spatially
with loops in our work, we are able to make a comparison with static loop scaling laws assuming the background field length is comparable to the length
of the loop embedded in it. We also discuss the implications of a breakdown in this assumption. Some very preliminary results from this
analysis appeared in \citet{winebarger_2012}.

\section{Data processing and methods}
\label{data}

\begin{figure*}
  \centerline{\includegraphics[viewport=90 250 520 740,clip,width=0.85\linewidth]{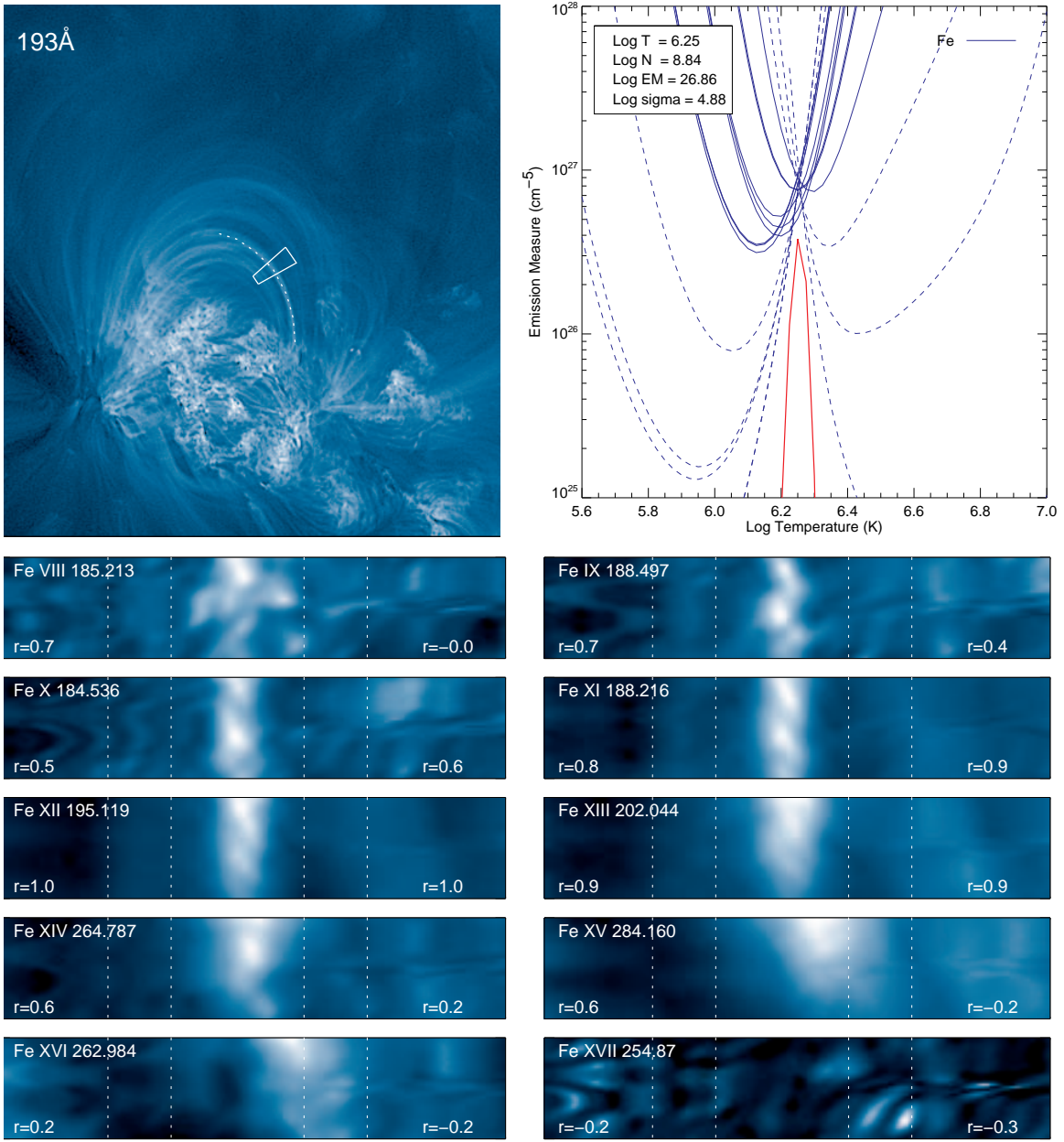}}
  \vspace{-0.1in}
\caption{ Images and emission measure distribution for the background around embedded loop \#8.
{\it Top Left}: AIA 193\,\AA\, image of the loop \#8 environment. The dotted line shows the trace used to
measure the loop half-length. The boxed region shows where the co-spatial background emission
was measured. {\it Top Right}: The EM distribution for the co-spatial background. The blue curves are
loci curves that indicate the constraints on the EM. Solid curves indicate spectral lines where
the background emission is correlated with the background emission in \ion{Fe}{12} 195.119\,\AA.
The dashed curves indicate spectral lines where the background emission is not correlated, and thus
the intensities were set to zero. The red curve is the Gaussian EM distribution. The derived physical
parameters are shown in the legend. {\it Lower Rows (2--6)}: Multi-wavelength images of the 
(straightened) embedded loop environment, also indicating the regions used for the cross-correlation 
analysis. The Pearson linear correlation coefficients ($r$) for the left and right regions are 
also shown.
\label{fig:fig1}}
\end{figure*}

\begin{figure*}
  \centerline{\includegraphics[viewport=90 50 520 740,clip,width=0.85\linewidth]{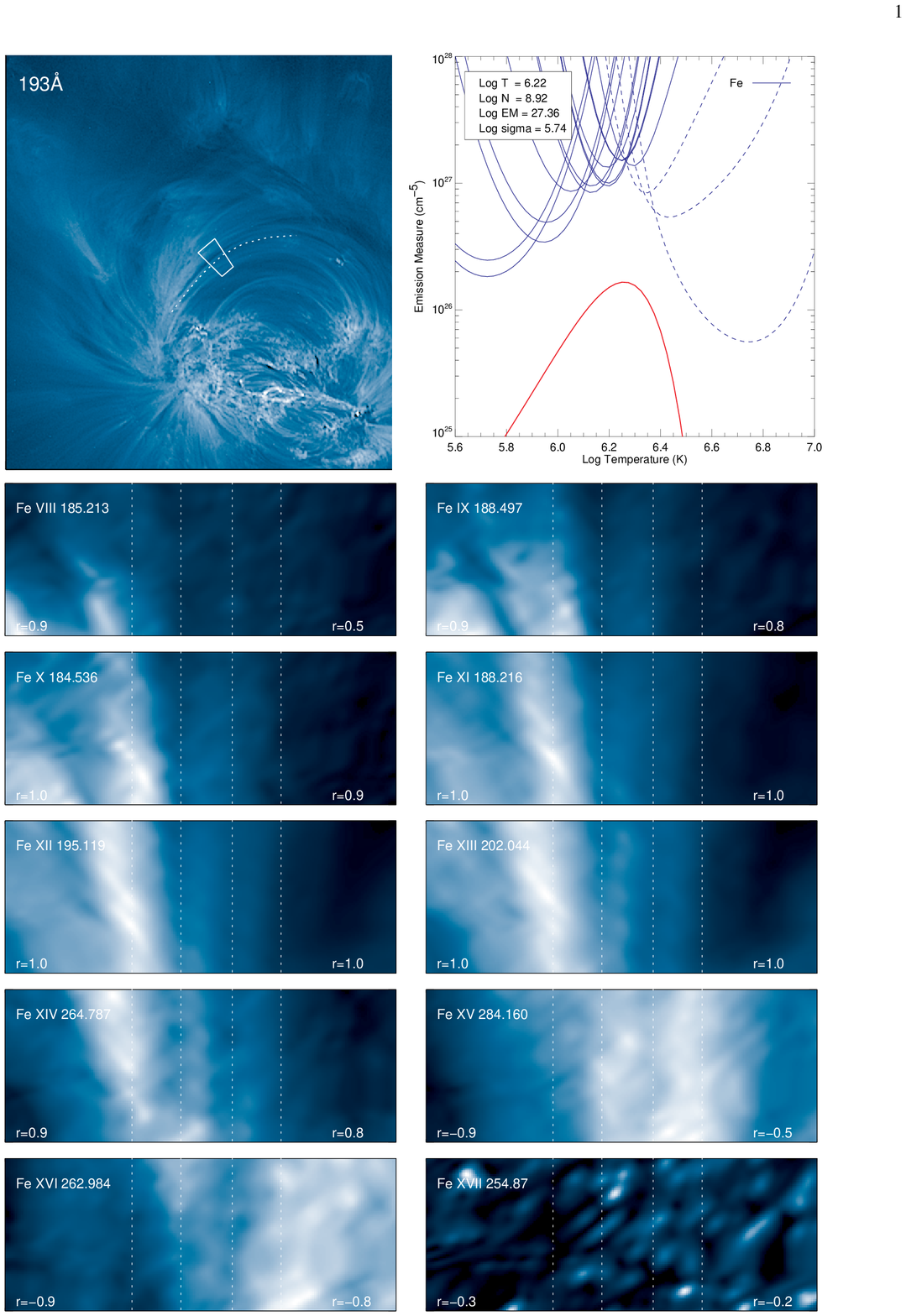}}
  \vspace{-0.1in}
\caption{ Same as Figure \ref{fig:fig1} but for the background around embedded loop \#6.
\label{fig:fig2}}
\end{figure*}

We use the sample of 20 warm loops that was previously analyzed by \citet{brooks_etal2012}, supplemented 
with 4 additional off-limb cases. Since line-of-sight superposition may be less at the limb, we wanted to
verify consistent behavior there, but no examples were included in the original sample. Using the same calibration
and analysis techniques as in \citet{brooks_etal2012}, we verified that the 
physical properties (temperature, density) of the new loops fall within the range of values found in the 
original sample.
The \citet{brooks_etal2012} sample come from EIS raster scans of several active regions present on the solar disk during 2011, 
and that study provides a detailed analysis of the physical properties of the loops, and comparisons between parameters
such as intensities, loop widths, and cross-loop intensity profiles derived from both the EIS data and observations from AIA.
The EIS observing sequence used the 1$''$ slit to scan an area of 240$''$ by 512$''$ in coarse
2$''$ steps. The sequence lasted about 2 hours and uses exposure times of 60s. The 4 additional examples 
were found after surveying data obtained using the same observing sequence through 2010--2012. We also used data 
from a narrower 120$''$ by 512$''$ scan that used 45s exposures and lasted around 45 mins. 
Both observing sequences recorded numerous strong 
spectral lines from the 180--285\,\AA\, wavelength range. The data were
then processed to remove instrumental effects, calibrated to physical units (erg cm$^{-2}$ s$^{-1}$ sr$^{-1}$)
and re-sampled to a 1$''$ scale. The SolarSoftware (SSW) routine eis\_prep was used for the data
processing. 

For this work, the spectroscopic analysis of emission measures, densities, and temperatures was 
performed on the EIS data. We also, however, use 1.7--4\,s exposure 193\,\AA\,
images from the AIA 
to measure the loop half-lengths. These 
data were processed for instrumental effects and made available online via the AIA cutout service. 
They are designated level 1.5. 

The EIS \ion{Fe}{12} 195.119\,\AA\, raster scan images were created by fitting the
spectra at each pixel to a double Gaussian function that takes account of both the main line and the 
weaker blend at 195.18\,\AA. Prior to extracting the background intensities we also fit single or double 
Gaussian functions to the spectra at
each pixel for all other wavelengths used in the analysis. For this study, we use the revised EIS 
intensity calibration of \citet{warren_etal2014}.

To measure the co-spatial background to each loop we examined the cross-field loop intensity profile.
We selected a clean portion of the loop, interpolated the intensities along the axis of the
loop, straightened the loop, and averaged the intensities along the loop segment. We then used the 
averaged cross-loop intensity profile to visually identify
two locations in the background close to the loop. We then fit the intensities between these locations  
with a first-order polynomial function. The measured background intensity is then the mean value of the 
polynomial function between the selected locations. The method is basically the same as established 
in \citet{aschwanden_etal2008} and \citet{warren_etal2008a} except that we extract the background information
rather than the loop information.
We show examples of the chosen loop segments on AIA 193\,\AA\, images in Figures \ref{fig:fig1} and \ref{fig:fig2}.

Also shown in Figures \ref{fig:fig1} and \ref{fig:fig2} are the traces used to measure the loop half-length. These
measurements were made using the same loop tracing software but in this case we only recorded the dimensions of
the straightened half-length segment. The loops were traced visually so there is some uncertainty in determining
the loop footpoints and apex and in accounting for any projection effects. 
We will discuss the influence of the loop length on our conclusions in Section \ref{results}.

We applied this technique to the EIS \ion{Fe}{12} 195.119\,\AA\, data and then extended it to the other spectral 
lines using the same background locations. We need spectral lines from a broad range of temperatures to
perform the EM analysis, so we used most of the strong lines from \ion{Fe}{8}--\ion{Fe}{17} that are observed by
EIS. These are listed in Table \ref{table1} and we have used most of them successfully in previous emission
measure studies \citep[see e.g.,][]{brooks&warren_2011}. At typical densities of the quiet Sun \citep[$\log$ (n/cm$^{-3}$) $=$ 8.5,][]{brooks_etal2009}
these lines cover the temperature range 0.52--5.5\,MK according to the CHIANTI v.8
database \citep{dere_etal1997,delzanna_etal2015}.

\begin{deluxetable}{ccc}
\tabletypesize{\scriptsize}
\tablewidth{0pt}
\tablecaption{EIS line-list used for the emission measure analysis. \label{table1}}
\tablehead{
\multicolumn{1}{c}{Element} &
\multicolumn{1}{c}{Ion} &
\multicolumn{1}{r}{Wavelength/\AA}
}
\startdata
Fe & VIII & 185.213 \\
Fe & VIII & 186.601 \\
Fe & IX   & 188.497 \\
Fe & IX   & 197.862 \\
Fe & X    & 184.536 \\
Fe & XI   & 188.216 \\
Fe & XI   & 188.299 \\
Fe & XI   & 192.813 \\
Fe & XII  & 186.880 \\
Fe & XII  & 192.394 \\
Fe & XII  & 195.119 \\
Fe & XIII & 202.044 \\
Fe & XIII & 203.826 \\
Fe & XIV  & 264.787 \\
Fe & XIV  & 270.519 \\
Fe & XV   & 284.160 \\
Fe & XVI  & 262.984 \\
Fe & XVII & 254.870
\enddata
\tablenotetext{*}{This is the complete list of lines used. They were not all used in the analysis of every dataset. }
\end{deluxetable}

We use CHIANTI v.8 to calculate all of the contribution functions we need for the EM analysis. The 
contribution function, $G(T,n)$, as a function of temperature ($T$) and density ($n$), is related to the 
observed line intensity for a transition between atomic levels $i$ and $j$ as
\begin{equation}
I_{ij} = \int G(T,n) \phi (T) dT
\end{equation}
where $\phi (T)$ is the differential emission measure re-cast as a function of temperature alone by assuming 
a fixed density at a given temperature. We use the \ion{Fe}{13} 202.044/203.826 diagnostic line ratio
to calculate the appropriate density in this work. We then fit the data with isothermal and Gaussian EM functions of the
form
\begin{equation}
\phi (T) = EM_0 \delta (T-T_0)
\label{eq2}
\end{equation}
and
\begin{equation}
\phi (T) = {EM_0 \over \sigma_T \sqrt {2 \pi} } \exp [- {(T-T_0) \over 2 \sigma^2_T} ]
\label{eq3}
\end{equation}
where $EM_0$ and $T_0$ are the peak emission measure and peak temperature, respectively, and $\sigma_T$ is the 
Gaussian width of the EM distribution. 
We also compute the reduced $\chi^2$ value for both models to allow a
comparison of the quality of the fit. 
We use Equation \ref{eq2} to test the isothermality of the plasma, and Equation \ref{eq3} to detect any
deviation from isothermal. The Gaussian EM has the advantage that if the plasma is multi-thermal, the width, $\sigma_T$, will increase, and
this provides a good measure of the degree of multi-thermality. A fixed Gaussian EM of course does not give us any information on the 
true shape of the EM distribution, but this is not a significant issue for our study.
We show examples of the EM analysis for two background regions in Figures \ref{fig:fig1} and \ref{fig:fig2}.
Figure \ref{fig:fig1} shows a case where the EM distribution is very close to isothermal, and Figure \ref{fig:fig2}
shows a case where the EM distribution is somewhat broader. 

For the EM analysis we also have to adopt some way to decide if the background emission at different wavelengths
comes from the same structure. In loop analysis we usually only include lines if the cross-field intensity
profile at that wavelength is highly correlated with the cross-field intensity profile at the wavelength used
to identify the loop. Here we looked at the spatial distribution of intensity in the straightened image segment,
parallel to the loop and within 5 pixels on either side. If the linear Pearson correlation coefficient, $r$, 
between the spatial distribution of intensity on either side of the loop and the spatial distribution of intensity
in the corresponding region of the \ion{Fe}{12} 195.119\,\AA\, image was greater than 0.75, then that 
wavelength was included. Figures \ref{fig:fig1} and \ref{fig:fig2} also show examples of the embedded loop environment
at different wavelengths 
and the background regions chosen for the cross-correlation analysis. We allow either side to be correlated 
because although we have tried to find clean
isolated loop segments, there is a natural transition in the structure of active regions as a function of
temperature that tends to favour excluding cooler lines on the inside of loop arcades and hotter lines on the outside. 
Figure \ref{fig:fig1} shows this effect for the cooler \ion{Fe}{8} 185.213\,\AA\, line. The \ion{Fe}{12} 195.119\,\AA\, 
emission on the outside of the loop in \ion{Fe}{8} 185.213\,\AA\, is correlated with the emission on the outside
of the \ion{Fe}{12} 195.119\,\AA\, loop ($r$=0.7), but the emission on the inside is uncorrelated ($r$=0.0) because
of the presence of hot core emission that is not seen at the temperatures of \ion{Fe}{8} 185.213\,\AA\,
but has an increasing contribution at the temperatures of \ion{Fe}{12} 195.119\,\AA.
Our method therefore may have a tendency to include more spectral lines in the analysis.
If the emission on neither side of the loop is highly correlated then the intensity is
set to zero and the error is set to 22\% 
of the background measurement
\citep[the approximate intensity calibration error, ][]{lang_etal2006}.

\begin{deluxetable*}{ccccccccccccccccccc}
\tabletypesize{\scriptsize}
\tablewidth{0pt}
\tablecaption{EIS measurements of the diffuse background emission. \label{table2}}
\tablehead{
\multicolumn{1}{c}{\#} &
\multicolumn{1}{c}{Date/Time} &
\multicolumn{1}{r}{$ELH$} &
\multicolumn{1}{c}{$\chi^2_i$} &
\multicolumn{1}{c}{$\chi^2_g$} &
\multicolumn{1}{c}{$T$} &
\multicolumn{1}{c}{$n$} &
\multicolumn{1}{c}{$EM$} &
\multicolumn{1}{c}{$\sigma_T$} &
\multicolumn{1}{c}{} &
\multicolumn{1}{c}{\#} &
\multicolumn{1}{c}{Date/Time} &
\multicolumn{1}{r}{$ELH$} &
\multicolumn{1}{c}{$\chi^2_i$} &
\multicolumn{1}{c}{$\chi^2_g$} &
\multicolumn{1}{c}{$T$} &
\multicolumn{1}{c}{$n$} &
\multicolumn{1}{c}{$EM$} &
\multicolumn{1}{c}{$\sigma_T$} 
}
\startdata
   1 & 01/21 13:57 & 20.9 & 1.05 & 1.15 & 6.22 & 8.99 & 26.8 & 4.50 & & 11 & 04/29 01:23 & 35.8 & 1.49 & 1.25 & 6.23 & 8.77 & 27.0 & 5.40 \\
   2 & 01/30 20:11 & 36.9 & 5.55 & 1.35 & 6.20 & 8.86 & 27.1 & 5.67 & & 12 & 05/06 13:55 & 19.4 & 2.60 & 1.16 & 6.31 & 9.03 & 27.2 & 5.72 \\
   3 & 03/18 09:34 & 62.6 & 0.63 & 0.69 & 6.25 & 8.70 & 26.9 & 4.50 & & 13 & 06/14 00:52 & 29.9 & 0.75 & 0.78 & 6.25 & 8.89 & 26.9 & 5.12 \\
   4 & 03/18 11:05 & 55.0 & 1.06 & 1.15 & 6.16 & 8.65 & 27.0 & 4.50 & & 14 & 07/02 03:32 & 28.0 & 0.78 & 0.86 & 6.27 & 8.88 & 26.9 & 4.50 \\
   5 & 04/15 00:41 & 61.4 & 4.27 & 1.93 & 6.36 & 8.99 & 27.6 & 5.92 & & 15 & 07/02 03:53 & 38.8 & 1.47 & 1.60 & 6.29 & 8.82 & 27.0 & 4.73 \\
   6 & 04/15 01:41 & 79.6 & 5.07 & 1.08 & 6.22 & 8.92 & 27.4 & 5.74 & & 16 & 09/02 13:25 & 68.9 & 0.78 & 0.84 & 6.25 & 8.85 & 27.1 & 4.88 \\
   7 & 04/21 12:02 & 25.6 & 1.16 & 0.88 & 6.31 & 8.74 & 26.9 & 5.49 & & 17 & 09/02 13:54 & 35.0 & 0.73 & 0.79 & 6.24 & 8.69 & 26.9 & 4.54 \\
   8 & 04/21 12:18 & 58.7 & 0.59 & 0.63 & 6.25 & 8.84 & 26.9 & 4.88 & & 18 & 09/17 18:57 & 44.6 & 8.67 & 1.38 & 6.50 & 9.03 & 27.6 & 6.12 \\
   9 & 04/22 08:56 & 44.0 & 1.26 & 0.85 & 6.32 & 8.86 & 27.3 & 5.54 & & 19 & 10/14 22:55 & 119.8 & 0.90 & 0.98 & 6.21 & 8.76 & 26.5 & 4.50 \\
   10 & 04/22 09:00 & 34.7 & 0.89 & 0.97 & 6.22 & 8.94 & 26.9 & 4.50 & & 20 & 10/14 23:51 & 24.6 & 4.09 & 2.56 & 6.35 & 8.74 & 27.4 & 5.68 \\
   -- & 10/09/05 04:10 & 124.1 & 2.65 & 1.13 & 6.20 & 8.46 & 27.1 & 5.45 & & -- & 06/27 12:08 & 80.7 & 7.15 & 0.58 & 6.24 & 8.79 & 27.6 & 5.75 \\
   -- & 06/27 12:14 & 80.0 & 0.75 & 0.82 & 6.31 & 8.63 & 27.4 & 4.55 & & -- & 12/05/16 23:52 & 68.5 & 1.95 & 1.27 & 6.23 & 8.64 & 27.2 & 5.42
\enddata
\tablenotetext{*}{The index indicates the loop number from \citet{brooks_etal2012} for comparison.
Additional new loops are unnumbered. All loops were observed in 2011 except where indicated. 
The EIS slit passed over the loop segment at the given date and time. $ELH$ is the half length of 
the loop embedded in the background emission. $\chi^2_i$ and $\chi^2_g$ are the reduced chi squared 
values for the isothermal and Gaussian models, respectively. $T$, $n$, $EM$, and $\sigma_T$ are the 
emission measure peak temperature, electron density, peak emission measure, and (Gaussian) emission 
measure width, respectively. The physical parameters are all from the Gaussian emission measure model. }
\end{deluxetable*}

\section{Results and discussion}
\label{results}

We give the results for the complete analysis of all 24 background regions in Table \ref{table2}.
The table shows the dates and times of all the observations we analyzed, together with a comparison of
the $\chi^2$ values for the isothermal and Gaussian EM analysis and the electron density evaluated 
from the \ion{Fe}{13} ratio. The $\chi^2$ for the
Gaussian model is closer to one in 88\% of the regions so we conclude that only three regions are likely
isothermal. Since the Gaussian model is a better fit in
general, we only report the peak temperature, peak EM, and thermal width from this model. Despite this,
about half the regions (11/24) have narrow thermal widths that are less than $\log$ (T/K) = 5 (7.6$\times$10$^{4}$ K).
Their peak temperatures fall in the range $\log$ (T/K) = 6.16--6.31 (1.4--2\,MK). Conversely, most of the
other regions (9/13) have thermal widths greater than $\log$ (T/K) = 5.5 (3$\times$10$^{5}$ K), and their
peak temperatures are $\log$ (T/K) = 6.2--6.5 (1.6--3.2\,MK). The densities for all the regions
are comparable. They fall in the range $\log$ (n/cm$^{-3}$) = 8.5--9.0.

The temperatures and densities for the background regions with narrow temperature distributions agree well 
with the previous results of \citet{delzanna&mason_2003} and \citet{subramanian_etal2014}, but the thermal 
width is much narrower, and hence the EM slopes are steeper. This is likely a result of only including
background emission that is correlated with the emission at the temperature of \ion{Fe}{12} 195.119\,\AA.

Regions with broader temperature distributions and higher peak temperatures (up to 3\,MK), however, have not
been discussed in this context before. Recall that this is diffuse emission around high lying warm loops, and it is not
clear if this is equivalent to the background emission around hot, compact core loops, though it seems likely
that this diffuse emission envelops the whole active region. That conjecture is supported by the work of
\citet{subramanian_etal2014}, who report that the diffuse emission above the active region core shows characteristics
similar to the hot core itself. 

\begin{figure}
  \centerline{\includegraphics[width=0.95\linewidth]{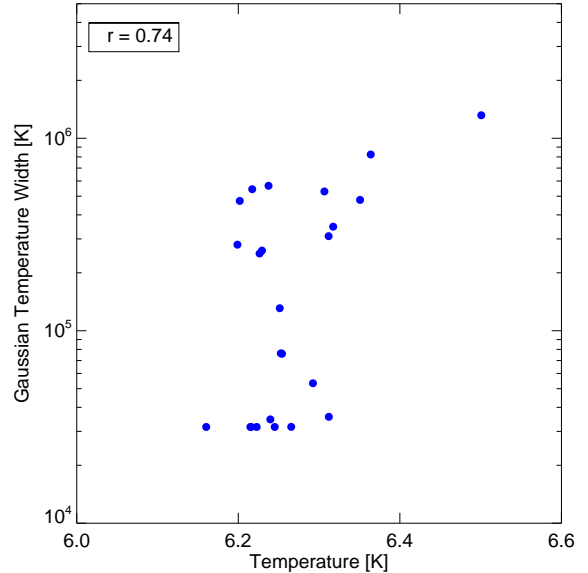}}
  \vspace{-0.1in}
\caption{ The relationship between peak temperature and the emission measure (Gaussian) width for the 
sample of diffuse background measurements. 
\label{fig:fig3}}
\end{figure}

Our results for the background emission around warm loops are also similar to the results for coronal loops.
Background emission peaked near 1.7\,MK\, has a narrow temperature distribution, but is generally not isothermal.
The higher temperature background emission has a broader temperature distribution. In fact, the width of 
the thermal distribution strongly correlates with the peak temperature. We plot the two quantities against
each other in Figure \ref{fig:fig3}. The linear Pearson correlation coefficient is $r$ = 0.74. This result is
similar to that reported for coronal loops by \citet{schmelz_etal2014}, who argued that the loop emission measure
distribution may narrow as it cools.

As discussed in the introduction, background subtraction is critical for measuring the properties of the loops themselves
\citep{klimchuk_etal1992,delzanna&mason_2003}. In this sample of loops, there is only one example where the background emission
is comparable in the two \ion{Fe}{13} lines used for measuring the electron densities. In the rest of the sample, not removing 
the background results in density changes of at least 40\%, rising to a maximum of a factor of 3.7. On average, densities would
be a factor of $\sim$2 lower if the background were not subtracted. Since the measurement is an average of the background and
loop intensity observed along the line of sight, this intuitively suggests that densities in the background emission should be
lower. This is exactly what we find in our analysis and is potentially significant. Densities in the background are
less than $\log$ (n/cm$^{-3}$) = 9, whereas warm loops generally have densities above 
$\log$ (n/cm$^{-3}$) = 9 \citep[see, e.g., ][]{brooks_etal2012}. 

These results suggest that the background plasma may be closer to static equilbrium. 
As discussed in Section \ref{introduction}, by measuring the background co-spatially with warm loops we have a
better measurement of the emission in which the loops are truly embedded. Therefore, we can make some assumption 
about the background field lengths based on the embedded loop length, and this allows us to make a direct comparison
with coronal loop scaling laws. As a first step in making some progress on the topic in this paper, we simply
assume that the background field length is the same as the measured loop length.
This allows us to calculate the column density, $nL$.

\begin{figure}
  \centerline{\includegraphics[width=0.95\linewidth]{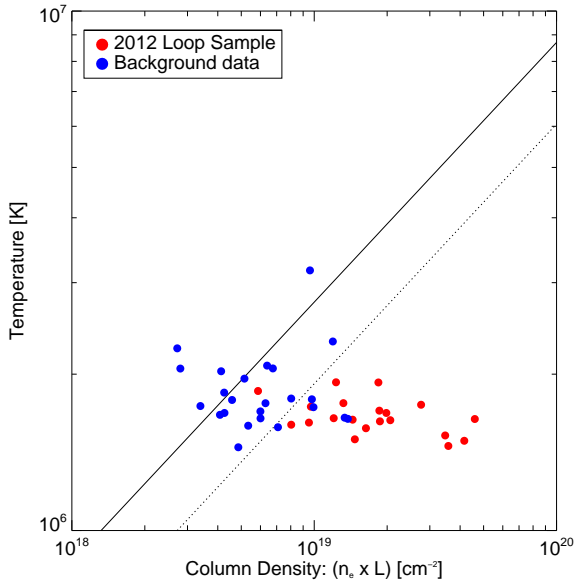}}
  \vspace{-0.1in}
\caption{ The relationship between measured temperature, $T$, and column density, ($n L$), for the 
diffuse background (blue) and a sample of coronal loops (red) from \citet{brooks_etal2012}. The solid
and dotted lines are theoretical scaling laws: 1.4$\times$10$^3$($pL$)$^{1/3}$ (RTV - solid) and 
1.1$\times$10$^3$($pL$)$^{1/3}$ (dotted).
\label{fig:fig4}}
\end{figure}

Figure \ref{fig:fig4} shows the relationship between the measured temperature and column density for 
the diffuse background data points (blue). We also show the corresponding values for the loops embedded in
the background (red). Also plotted on the graph are two theoretical scaling laws. The solid line is the well-known
RTV law \citep{rosner_etal1978}. Strictly speaking this scaling relates the maximum temperature 
of the loop to the column density assuming the heat source is located at the loop top. The dotted line
is the scaling law assuming
that the heating occurs uniformly along the loop \citep{kano&tsuneta_1995}. Note that in our loop sample
not all of our measurements
are made at the loop top, and may also not be located at the loop maximum temperature. The locations were
chosen to get a clean cross-field intensity profile for a different purpose. 
We only show these data points, for comparison, however. With no discernible structure, it is not possible
to verify the top of the background emission.

The Figure shows the well known result that warm loops are over-dense quite clearly. While a few of the warm loops
might be said to agree with the uniform heating scaling law, the mean 
loop density is $\log$ (n/cm$^{-3}$) = 9.4 at the mean temperature of 1.7\,MK, and this is about a factor of 5.4 larger than
the expected density from the RTV scaling law, and a factor of 2.6 larger than the expected density from
the uniform heating scaling law. In contrast, most of the background emission appears broadly consistent with
one or other of the scaling laws. The densities of 10/24 regions are within 40\% of the expected values from
the RTV scaling law, and 10 others are within 40\% of the expected values from the uniform heating law. Only 
2/24 are under-dense compared to the RTV scaling prediction - again similar to some
reports of densities in the high temperature core emission \citep{winebarger_etal2003b,klimchuk_2006} - and two others are over-dense compared
to the uniform heating law. 

We recognize that our simple assumption about the field line length is an approximation that affects these results.
There are a few background regions where the agreement with static theory would begin to break down if the field 
lines were $\sim$30--50\% longer.
Since the mean background density is $\log$ (n/cm$^{-3}$) = 8.8 and the mean temperature is 1.9\,MK, however, 
the field line lengths would have to be factors of 2--3 longer on average to produce a similar disagreement with theory to that 
which is seen for the warm loops. We do not expect that our loop length measurements are inaccurate to this
degree, but of course we cannot rule out some unexpected property of the background field that means it does
not track the loop closely, e.g., if it reaches much higher altitudes near the apex, or if the eccentricity is
lower than for the loop and the separation between footpoints is much greater. These factors, however, lead to 
several cases where the loop length approaches the largest in our sample (half-length of 120\,Mm), and, assuming
a circular loop, footpoint separations of $>$200$''$. These dimensions are typical only of the largest active 
regions.

Conversely, we know that the emission we measure is an integrated total along the line-of-sight, and the
path length is likely to be quite large. Estimates assuming an isothermal plasma and an exponential fall-off 
in density with a hydrostatic scale-height suggest that the path length needs to be 160-200\,Mm to 
encompass 80\% of the emission along the line-of-sight \citep{aschwanden&acton_2001,warren&warshall_2002,brooks&warren_2008}. 
This will of course include emission from both large-scale overlying foreground magnetic field, and small-scale
background field. It would be interesting to model the background magnetic field and determine whether the 
integrated line-of-sight intensity is dominated by emission from an ensemble of field lines with an average length 
close to that of the embedded loop. While at first sight this may appear unlikely, it also seems unlikely that 
it would be dominated by only the largest scale magnetic field because the emission is dropping with
height, and this is what is needed to break the agreement with the static loop scaling laws (based on the simple
assumption that the field length is comparable to the embedded loop length).

In summary, we have studied the densities and temperatures of a sample of 24 diffuse background/foreground emission regions that are co-spatial
with embedded warm loops, and where the emission at different temperatures comes from the same structures (based on
our cross-correlation analysis). By analyzing the co-spatial background we obtain results that are more representative
of the embedded loop environment, and we are also able to use the loop information to guide assumptions about the length of the background
field. In this way we were able to compare the background properties to expectations from static loop scaling laws. The
results we present here suggest that the properties of the 
majority of the background regions in our sample are compatible with quasi-steady, or high frequency impulsive heating.


\acknowledgments 

DHB would like to thank Harry Warren for helpful comments on the manuscript.
The work of DHB was performed under contract to the Naval Research Laboratory and was funded
by the NASA \textit{Hinode} program. \textit{Hinode} is a Japanese mission developed and launched 
by ISAS/JAXA, collaborating with NAOJ as a domestic partner, NASA and STFC (UK) as international 
partners. Scientific operation of the \textit{Hinode} mission is conducted by the \textit{Hinode} 
science team organized at ISAS/JAXA. This team mainly consists of scientists from institutes in 
the partner countries. Support for the post-launch operation is provided by JAXA and NAOJ(Japan), 
STFC (U.K.), NASA, ESA, and NSC (Norway). Courtesy of the NASA/SDO, and the AIA, EVE, and HMI
science teams. CHIANTI is a collaborative project involving George Mason University, the University 
of Michigan (USA) and the University of Cambridge (UK).


\end{document}